\documentclass[11pt]{article}
\usepackage[textwidth=15.2cm,textheight=22cm]{geometry}
\usepackage{amsmath,amssymb}
\usepackage{latexsym}
\usepackage{graphicx}
\usepackage{bbm}
\usepackage{bm}
\usepackage{cite}
\allowdisplaybreaks[1]
\newcommand{\be}{\begin{equation}}
\newcommand{\ee}{\end{equation}}

\newcommand\fr[1]{\frac{1}{#1}}
\def\k{\kappa}

\newcommand{\nn}{\nonumber}

\def\bea{\begin{eqnarray}}
\def\eea{\end{eqnarray}}
\def\beas{\begin{eqnarray*}}
\def\eeas{\end{eqnarray*}}
\def\ndt{\noindent}
\def\sla{\raise.15ex\hbox{$/$}\kern-.57em}

\def\spa#1.#2{\left\langle#1\,#2\right\rangle}
\def\spb#1.#2{\left[#1\,#2\right]}

\begin{document}

\begin{titlepage}
\vskip 1cm
\centerline{\Large{\bf {Gravity and Yang-Mills theory}}} 
\vskip 1.5cm
\centerline{Sudarshan Ananth}
\vskip .5cm
\centerline{\it {Indian Institute of Science Education and Research}}
\centerline{\it {Pune 411021, India}}
\vskip 1.5cm
\centerline{\bf {Abstract}}
\vskip .5cm
\ndt Three of the four forces of Nature are described by quantum Yang-Mills theories with remarkable precision. The fourth force, gravity, is described classically by the Einstein-Hilbert theory. There appears to be an inherent incompatibility between quantum mechanics and the Einstein-Hilbert theory which prevents us from developing a consistent quantum theory of gravity. The Einstein-Hilbert theory is therefore believed to differ greatly from Yang-Mills theory (which does have a sensible quantum mechanical description). It is therefore very surprising that these two theories actually share close perturbative ties. This article focuses on these ties between Yang-Mills theory and the Einstein-Hilbert theory. We discuss the origin of these ties and their implications for a quantum theory of gravity.
\vfill
\end{titlepage}

\ndt Gravity, although the oldest force known to man, is the force we understand the least. The weakness of the gravitational force precludes the possibility of performing simple experiments that will teach us more about this omnipresent force. In addition, the fact that the gravitational constant $\k$ is dimensionful seems to produce a plethora of problems when attempting to unite quantum mechanics and general relativity.

\vskip 0.2cm

\ndt Apart from gravity there are three other {\it {fundamental}} forces in Nature: the electromagnetic force, the weak force and the strong force. These three forces are all described by quantum Yang-Mills theories and theoretical predictions have, with stunning accuracy, matched experimental observations. This leads us naturally back to gravity: to wonder why, unlike the other three forces, it refuses to play well in the quantum mechanical playground. Experimental checks on the general theory of relativity have proved very succesful and this suggests that the correct quantum theory of gravity will prove to be an ``extension" of Einstein's theory rather than a replacement. 

\vskip 0.2cm

\ndt There are many reasons to believe that gravity is indeed very different from the other three forces. The Einstein-Hilbert Lagrangian, which describes gravity, differs greatly from the Yang-Mills Lagrangian suggesting that the theories they describe should also differ drastically. 

\vskip 0.2cm

\ndt In this article we paint a picture that conveys quite the opposite impression - a picture where Yang-Mills and gravity seem to have far more in common than previously believed. We explore close ties between the theories that go beyond superficial on-shell relations.

\vskip 0.2cm

\ndt We begin with a quick review of why the two theories are expected to behave very differently. The Yang-Mills action reads
\bea
{\it S}_{\mbox {\tiny {YM}}}\,=\,-\fr{4}\int d^4x\;{\mbox {Tr}}\,(F^{\mu\nu}F_{\mu\nu})\ ,
\eea
with
\bea
F_{\mu\nu}=\partial_\mu A_\nu-\partial_\nu A_\mu+i\,g\,[A_\mu,A_\nu]\ .
\eea
$A_\mu$ represents the gauge field and $g$ the dimensionless Yang-Mills coupling constant. Gravity, on the other hand, is governed by the Einstein-Hilbert action
\bea
{\it S}_{\mbox {\tiny {EH}}}\,=\,\fr{\k^2}\,\int d^4x\,\sqrt {-{\mbox {g}}}\,R\ ,
\eea
where $R$ is the Ricci scalar, ${\mbox {g}}$ the determinant of the metric and $\k$ the dimensionful gravitational coupling constant. Significant differences between the two theories include
\begin{enumerate}
\item Yang-Mills theory has only cubic and quartic interaction vertices while gravity involves infinitely many interaction vertices.
\item The coupling constant $g$ is dimensionless while $\k$ has the dimensions of length.
\item The trace in Yang-Mills, due to the gauge group, is absent in gravity.
\end{enumerate}
Each of these is a fairly substantial difference in its own right. Given these manifest differences, close ties between the two theories are all the more surprising. 

\vskip 0.2cm

\ndt The first clear indications of a concrete connection between gravity and Yang-Mills theory arose from the Kawai-Lewellyn-Tye (KLT) relations~\cite{KLT}. These relations tells us that tree-level scattering amplitudes in pure gravity are the ``square" of tree-level scattering amplitudes in pure Yang-Mills theory. Thus, rather unexpectedly, graviton scattering is expressible as a sum of products of pieces of non-abelian gauge theory scattering amplitudes~\cite{ZB}. Since scattering amplitudes contain information about the physics of a process, this points to a much stronger relationship between gravity and Yang-Mills theory than one might anticipate from a cursory glance at the respective Lagrangians.

\vskip 0.2cm

\ndt The KLT relations are easy to write down but before doing so, we need to address the differences, between the two theories, listed above in points 2 and 3. To this end we define, for Yang-Mills, color-stripped partial amplitudes
\bea
{\it {A}}^{\rm {tree}}_n=g^{(n-2)}\,{\mbox {Tr}}(\ldots)\,\times\,{\mbox {A}}^{\rm {tree}}_n\ .
\eea
${\it {A}}^{\rm {tree}}_n$ is a tree-level scattering process involving $n$ gluons in pure Yang-Mills theory. All information regarding the color structure is contained within the trace and ${\mbox {A}}^{\rm {tree}}_n$ now represents the color-stripped partial amplitude. Similarly, in the case of gravity, we define
\bea
{\it {M}}^{\rm {tree}}_n={{\bigg (}\frac{\kappa}{2}{\biggr )}}^{(n-2)}\,\times\,{\mbox {M}}^{\rm {tree}}_n\ ,
\eea
where ${\it {M}}^{\rm {tree}}_n$ represents a tree-level gravity scattering process and $\k^2=32\pi G_N$ is the coupling in terms of the Newton constant. ${\mbox {M}}^{\rm {tree}}_n$ is now $\k$-independent and represents the coupling-stripped gravity amplitude. We are now in a position to write down the KLT relations
\bea
{\mbox {M}}^{\rm {tree}}_n\propto ({\mbox {A}}^{\rm {tree}}_n)^2\ ,
\eea
suggesting the possibility of a much broader and intriguing relation of the form
\bea
\label{genrel}
{\mbox {Gravity}} \sim ({\mbox {Yang-Mills}}) \times ({\mbox {Yang-Mills}})\ .
\eea
Such a broad relationship however would require far more than simple tree-level connections. If the relationship is indeed deeper it ought to stem from similarities at the off-shell level: in particular, {\it {can we relate the two theories directly at the level of the Lagrangians?}} It turns out that the answer to this question is at least partially a yes and the rest of this paper will expand on this point.

\vskip 0.2cm

\ndt We start by observing that {\it {both}} the gauge field and the graviton field, in four dimensions, involve exactly two physical degrees of freedom\footnote{These degrees of freedom correspond to the positive and negative helicity states under the $SO(2)$ little group in four dimensions.}. We make this degrees-of-freedom equality manifest in the Lagrangians by working in light-cone gauge where only physical helicity states are present. 

\vskip 0.2cm

\ndt The structure of the light-cone gauge Yang-Mills Lagrangian, in momentum space, is
\bea
\label{schym}
{\it L}_{\mbox {\tiny {YM}}}\,\sim\,\bar A\,p_\mu^2\, A+g\,f_1\,\bar AAA+g\,f_2\,\bar A\bar A A+g^2\,f_3\,\bar A\bar A AA\ ,
\eea
where $\bar A$, $A$ are $-,+$ helicity states of the gauge field, $p_\mu^2$ stems from the d'Alembertian in the kinetic term, $g$ is the Yang-Mills coupling constant and $f_1,f_2,f_3$ represent momentum coefficients\footnote{Integrals over momenta and $\delta$-functions are not shown here explicitly.}. 

\vskip 0.2cm

\ndt It turns out, from explicit calculations, that tree-level scattering amplitudes constructed using the $\bar AAA$ (helicity {\tiny {$-++$}}) vertex in (\ref {schym}) vanish~\cite{LD}. So, when focusing on amplitude calculations it makes sense to try and eliminate this vertex by means of a field redefinition. Such a canonical change-of-variables, for Yang-Mills theory, was found in~\cite{GR,PM}. The change-of-variables maps the first two terms of the interacting Yang-Mills Lagrangian in (\ref {schym}) into a free Lagrangian in new variables
\beas
\bar A\,p_\mu^2\, A+g\,f_1\,\bar AAA\,\rightarrow\,\bar B\,p_\mu^2\, B\ .
\eeas
This is achieved by the following canonical redefinition of the gauge field~\cite{GR,PM}
\bea
\begin{split}
A&\,\rightarrow\,B+b_1B^2+b_2B^3+\ldots+b_{n-1}B^n \ , \nn \\
\bar A&\,\rightarrow\,\bar B+c_1B\bar B+c_2B^2\bar B+\ldots+c_{n-1}B^{n-1}\bar B\ ,
\end{split}
\eea
where $b_n, c_n$ are functions of the momenta and $\bar B, B$ represent the ``shifted" gauge field. This canonical change-of-variables results in a new form for the Yang-Mills Lagrangian
\bea
\label{schnym}
{\it L}^\prime_{\mbox {\tiny {YM}}}\,\sim\,\bar B\,p_\mu^2\, B+g\,F_1\,\bar B\bar B B+g^2\,F_2\,\bar B\bar B BB+\ldots+g^{n-1}\,F_{n-1}\,\bar B\bar B B^n\ ,
\eea
where the $F_n$ are the mometum coefficients for the new interaction vertices. The price paid for eliminating the offending term from (\ref {schym}) is the appearance of infinitely many interaction vertices {\it {exactly like in gravity}}. Note that all interaction vertices in (\ref {schnym}) involve exactly two negative helicity fields\footnote{The apparent disparity in the Lagrangian between $+$ and $-$ is an artifact of the convention-choice~\cite{LD}.}. 

\vskip 0.2cm

\ndt The advantage of the new form of the Yang-Mills Lagrangian is that certain classes of scattering amplitudes (referred to as MHV) become trivial to compute. For example, the amplitude for {\it {A}}{\tiny {$(--++)$}} is obtained trivially by taking the coefficient $F_2$ from (\ref {schnym}) and putting it on-shell. To compute the same process starting from (\ref {schym}), one would have to deal with not just the quartic vertex but also contact diagrams arising from combinations of the two cubic vertices.

\vskip 0.2cm

\ndt We now have a ``close to on-shell physics" form for the Yang-Mills Lagrangian where amplitude structures are manifest. Since amplitudes in Yang-Mills are related to those in gravity, it seems natural to look for a similar amplitude-friendly form for the gravity Lagrangian - such a gravity Lagrangian ought to have manifest similarities to (\ref {schnym}). 

\vskip 0.2cm

\ndt We start with the momentum-space Einstein-Hilbert Lagrangian in light-cone gauge~\cite{SS} 
\bea
\label{schg}
{\it L}_{\mbox {\tiny {EH}}}\,\sim\,\bar h\,p_\mu^2\, h+\kappa\,l_1\,\bar hhh+\kappa\,l_2\,\bar h\bar h h+\kappa^2\,l_3\,\bar h\bar h hh+\ldots
\eea
where $\bar h$, $h$ are $-,+$ helicity states of the graviton and $l_1,l_2,l_3,\ldots$ represent momentum coefficients. Notice that this structure, which is similar in some ways to (\ref {schym}), reveals the key difference highlighted earlier: that Yang-Mills theory involves only a finite number of interaction vertices (making it renormalizable) while gravity involves infinitely many interaction vertices. As with Yang-Mills, amplitudes arising from the first cubic vertex in (\ref {schg}) vanish motivating the elimination of this vertex.

\vskip 0.2cm

\ndt A canonical field redefintion, analagous to the one in Yang-Mills, was found in~\cite{AT}
\bea
\begin{split}
h\,&\rightarrow\,C+q_1C^2+q_2C^3+\ldots+q_{n-1}C^n\ , \nn \\
\bar h\,&\rightarrow\,\bar C+r_1C\bar C+r_2C^2\bar C+\ldots+r_{n-1}C^{n-1}\bar C\ ,
\end{split}
\eea
where $q_n, r_n$ represent functions of the momenta and $\bar C, C$ the ``shifted" graviton. This redefinition sucessfully eliminates the ``bad" vertex resulting in the new form
\bea
\label{schng}
{\it L}^\prime_{\mbox {\tiny {EH}}}\,\sim\,\bar C\,p_\mu^2\, C+\kappa\,L_1\,\bar C\bar C C+\kappa^2\,L_2\,\bar C\bar C CC+\ldots+\kappa^{n-1}\,L_{n-1}\,\bar C\cdots\bar C\,C\cdots C+\ldots
\eea
where the $L_n$ are the momentum coefficients for the new interaction vertices. 

\vskip 0.2cm

We are now in a position to compare the two structures in (\ref {schnym}) and (\ref {schng}). Up to second order in the coupling constants we find that gravity does indeed behave like the ``square" of Yang-Mills. For instance
\bea
\label{klt}
L_2\,\sim\,(F_2)\times(F_2)\ .
\eea
This is a much stronger statement than a KLT relation since it relates {\it {off-shell}} coefficients in a Lagrangian as opposed to on-shell amplitudes. Unfortunately, beyond second order in the coupling new problems crop up. This happens because gravity vertices in (\ref {schng}) involve varying numbers of fields of both helicities, unlike Yang-Mills where all vertices in (\ref {schnym}) involve exactly two negative helicity fields. A complete understanding of the relationship in (\ref {genrel}) could prove invaluable for a detailed finiteness-analysis~\cite{AKS} of gravity or supergravity. 

\vskip 1 cm
\begin{center}
* ~ * ~ *
\end{center}
\vskip 0.5 cm

\ndt In this article, we have reviewed two equivalent forms of the Yang-Mills Lagrangian. The first has a finite number of interaction vertices while the second involves infinitely many such vertices. We have also described two equivalent light-cone Lagrangians for gravity, both involving infinitely many vertices. We conclude by asking whether there exists a third equivalent form of the gravity Lagrangian that involves only a finite number of interaction vertices\footnote{The answer to this question is most likely a No. However, where the search for a quantum theory of gravity is concerned, concrete avenues of progress are rare and this approach, even if unsuccesful, is certain to teach us more about the UV structure of gravity - this makes it worth pursuing to its logical conclusion.}. If yes, what implications will such a form of the Lagrangian have for the ultra-violet properties of gravity? 

\vskip 0.5cm

\ndt {\it {Acknowledgments}}: This work was supported by a Ramanujan Fellowship from the Department of Science and Technology (DST), Government of India and by the Max Planck Society and DST through the Max Planck Partner Group in Quantum Field Theory.

\end{document}